# Towards On-Chip MEMS-Based Optical Autocorrelator

Ahmed M. Othman, *Student Member, IEEE*, Hussein E. Kotb, *Member, IEEE*, Yasser M. Sabry, *Member, IEEE*, Osama Terra, and Diaa A. Khalil, *Senior Member, IEEE*

*Abstract*—We propose a compact MEMS-based optical autocorrelator based on a micromachined Michelson interferometer in silicon and the two-photon absorption non-linearity in a photodetector. The miniaturized autocorrelator has a scanning range of 1.2 ps and operates in the wavelength range of 1100-2000 nm. The device measures the interferometric autocorrelation due to its collinear nature, from which the intensity autocorrelation can be calculated. The field autocorrelation can also be measured, from which the optical pulse spectrum can be calculated. A theoretical model based on Gaussian beam propagation is developed to study the effect of optical beam divergence, pulse dispersion, tilt angle between the interferometer mirrors, and amplitude mismatch between the interfering pulses. This model explains many of the effects observed in experimental measurements due to the use of a MEMS interferometer. The experimental results of autocorrelation signals for several pulses in the order of 100 fs are compared to a commercial autocorrelator and a good match is found.

## I. Introduction

OPTICAL autocorrelators are used for measuring ultrashort pulses that have a width in the order of tens of picoseconds or less. These pulses cannot be measured directly using conventional photodetectors due to the slow response time of the latter. Thus, autocorrelators are useful in the development of ultrashort pulsed sources such as mode-locked lasers, supercontinuum laser sources and optical frequency combs. These sources have a wide range of applications from spectroscopy and optical communication to applications in the biomedical domain [1]–[3]. However, autocorrelators are usually bulky and expensive due to the use of many components and mechanical moving parts that require precise alignment.

Many efforts have been exerted recently to design a compact autocorrelator based on silicon photonics. But the reported devices are either incapable of measuring sub-ps pulses [4]–[6], have a very limited wavelength range of operation [7], or only capable of measuring pulses having a time-bandwidth product greater than 100 [8]. In addition, some work has been reported based on CdS or CdTe nanowires [9]–[10] but the use of alignment-sensitive optical components for coupling light into the nanowires is still a challenge. Another technique that requires high spatial coherence for few-cycle pulses measurement using an angular tunable bi-mirror for non-collinear autocorrelation is reported in ref. [11].

In this work, a MEMS-based autocorrelator that uses a Michelson interferometer fabricated using silicon micromachining technology is reported. The device uses the two-photon absorption (TPA) non-linearity in a silicon detector, allowing the potential of integration into a single chip. The rest of this paper is organized as follows. Section II reviews the background of optical autocorrelation and describes the MEMS device. The experimental results are presented in section III. Section IV discusses the non-ideal effects that can be present in the interferometer such as beam divergence, silicon dispersion, non-vertical mirror surfaces and amplitude mismatch between the interfering pulses. Finally, the work is concluded in section V.

## II. Theoretical Background and Device Description

A typical collinear autocorrelator uses a Michelson interferometer as shown in Fig. 1, where the input pulse is split into two pulses using a beam splitter. One of the two interferometer arms has a moving mirror to allow scanning the delay $\tau$ between the two interfering beams. The output electric field from the interferometer in the time domain can then be written as [12]:

$$E_{out}(t;\tau) = \tfrac{1}{2}\,\text{Re}\{(E_p(t) + E_p(t-\tau)e^{-j\omega_0\tau})e^{j\omega_0 t}\} \qquad (1)$$

where $E_p(t)$ is the temporal pulse shape and $\omega_0$ is the optical angular frequency of the pulse. Using a slow linear photodetector at the output of the interferometer, the output

A. M. Othman is with the Faculty of Engineering, Ain Shams University, Cairo 11517, Egypt (e-mail: ahmed.othman@eng.asu.edu.eg).
H. E. Kotb is with the Transmission Department, National Telecommunication Institute, Cairo 11768, Egypt (e-mail: hussein.kotb@nti.sci.eg).
Y. M. Sabry and D. A. Khalil are with the Faculty of Engineering, Ain Shams University, Cairo 11517, Egypt. They are also with Si-Ware Systems, Cairo 11361, Egypt (e-mail: yasser.sabry@eng.asu.edu.eg; diaa.khalil@si-ware.com).
O. Terra is with the Primary Length Standard and Laser Technology Laboratory, National Institute of Standard, Giza 12211, Egypt (e-mail: osama.terra@nis.sci.eg; osama.terra@gmail.com).

current from the detector yields the field autocorrelation signal

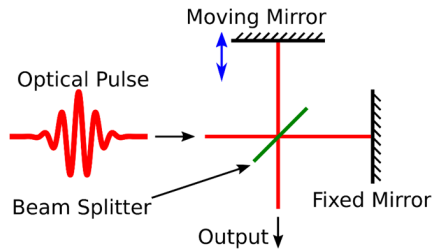

Fig. 1. Schematic of a Michelson interferometer.

given by:

$$I_{field}(\tau) = C_1 \left\{ \int \left( |E_p(t)|^2 + [E_p(t)E_p^*(t-\tau)e^{j\omega_0\tau} + c.c.] \right) dt \right\} \quad (2)$$

where $I_{field}(\tau)$ is the detector current versus delay, $C_1$ is a constant and c.c. denotes the complex conjugate. The Fourier transform of the field autocorrelation signal yields the optical spectrum of the input pulse [13].

Adding a non-linear element before the detector, the output of the detector versus delay yields the interferometric autocorrelation signal. The output current from the detector in this case is written as [11]:

$$I_{interf}(\tau) = 2C_2 \int \left( |E_p(t)|^4 + 2|E_p(t)|^2|E_p(t-\tau)|^2 \right) dt + C_2 \left\{ 2e^{j\omega_0\tau} \int E_p(t)E_p^*(t-\tau) \left[ |E_p(t)|^2 + |E_p(t-\tau)|^2 \right] dt + e^{j2\omega_0\tau} \int [E_p(t)E_p^*(t-\tau)]^2 dt + c.c. \right\} \quad (3)$$

where $I_{interf}(\tau)$ is the detector current for the interferometric autocorrelation signal, and $C_2$ is a constant. The exponential terms in this equation are fast-varying terms, which can be suppressed by averaging the interferometric autocorrelation signal over many fringes, yielding the intensity autocorrelation signal:

$$I_{inten}(\tau) = 2C_2 \int \left( |E_p(t)|^4 + 2|E_p(t)|^2|E_p(t-\tau)|^2 \right) dt \quad (4)$$

which can be used to get the pulse width of the input pulse. It is worth mentioning that the autocorrelation signal, defined by (3) or (4), is ideally an even function of $\tau$ regardless of the symmetry of the actual pulse. Furthermore, the ratio between the maximum value and the background level for the interferometric and intensity autocorrelation signals is 8 to 1 and 3 to 1, respectively [11].

The proposed device uses a MEMS interferometer, as shown in Fig. 2(a). The MEMS interferometer is composed of a silicon beam splitter, a fixed mirror and a moving mirror driven by a comb-drive actuator as shown in Fig. 2(b). All the components of the MEMS chip are fabricated at the same time in a self-aligned manner, which is crucial for the device operation. The self-alignment is enabled by the photolithographic accuracy and subsequent etching [14]–[19]. The mirror metallization is achieved using step coverage of vertical surfaces [20]. Light is propagating in-plane with respect to the chip substrate in a micro-optical bench arrangement. The output beam from the MEMS interferometer is then focused on a silicon detector that is used outside its linear absorption wavelength range to exploit the TPA process as a source for the nonlinearity necessary for obtaining information about the pulse width. The output current from the detector is measured and the interferometric autocorrelation signal can be constructed. The silicon detector can also be replaced in the same setup by an InGaAs photodetector, allowing the field autocorrelation signal to be measured. The Fourier transform of the field autocorrelation signal yields the pulse optical spectrum. Therefore, the presented device can be used as an optical autocorrelator or a spectrometer for full characterization of the pulse in the time and wavelength domains.

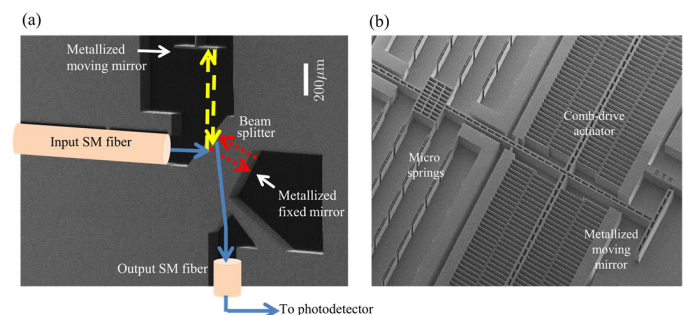

Fig. 2. SEM image for a fabricated MEMS-based (a) Michelson interferometer and (b) comb-drive actuator.

## III. EXPERIMENTAL RESULTS

### A. Device and Measurement Setup

A schematic of the proposed autocorrelator and the measurement setup is shown in Fig. 3. The input pulse under test is fed to the MEMS interferometer by the means of a GRIN lensed fiber that partially collimates light to $W_0$ of about 10 $\mu m$ to decrease the divergence losses inside the interferometer. The output beam from the interferometer is then tightly focused on a silicon avalanche photodetector (Thorlabs APD130A2) using a microscope objective lens with a negligible dispersion effect on the measured pulse width. The silicon photodetector generates a current proportional to the square of the input optical intensity by the TPA process; allowing it to replace the second harmonic generation (SHG) crystal typically used in scanning autocorrelators [21]. Since the MEMS interferometer is also fabricated in silicon, the proposed autocorrelator has the potential of integration into a single chip by combining the photodetector onto the same die. For the TPA to be the dominant absorption mechanism, the input pulse wavelength should be outside the linear absorption range of silicon (400 nm – 1100 nm). In addition, the input pulse wavelength should be within double the wavelength range of the linear absorption of silicon. Combining these two conditions determines the possible wavelength range of the device to be from 1100 nm to about 2000 nm. To measure the pulse autocorrelation signal, electronic circuits drive the MEMS-based comb-drive actuator and measure the current from the photodetector. The measured signals are then processed to obtain the pulse autocorrelation signal versus the temporal delay.

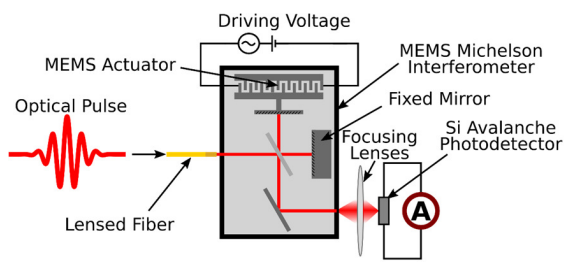

Fig. 3. Schematic of the proposed autocorrelator and the experimental setup for autocorrelation measurement.

## B. Measurement Results

To test the autocorrelator, a femtosecond mode-locked fiber laser is fed to the autocorrelator input to measure its pulse width. The mode-locked laser has a central wavelength and a pulse repetition rate of 1560 nm and 16 MHz, respectively. The gain medium of the laser is an erbium doped fiber that is pumped by a laser diode at 980 nm. The output pulse shape and width from the source can be changed by changing the pump laser diode output power. The pulse interferometric autocorrelation was measured at 3 values of pump laser diode power; namely 180 mW, 215 mW and 240 mW as shown in Fig. 4. The fringe-averaged intensity autocorrelation was calculated from the interferometric autocorrelation and the result was compared to the intensity autocorrelation measured using a commercial autocorrelator (A. P. E. pulseCheck). The measured autocorrelation full width half maximum (FWHM) using the proposed device is 178 fs, 183 fs and 166 fs corresponding to pulse widths of 126 fs, 130 fs and 117 fs, respectively, assuming a Gaussian pulse shape for the deconvolution factor. Table 1 lists the measured pulse width at different values of mode-locked laser pump power for the proposed MEMS-based autocorrelator and the commercial device, where the results indicate a good match.

The proposed device is also used to measure the pulse optical spectrum only by replacing the silicon detector by an InGaAs detector, which works as a linear detector. The measured autocorrelation signal in this case is the field autocorrelation. The Fourier transform of the field autocorrelation yields the pulse optical power spectrum. The proposed device may be potentially used to measure the pulse width and optical spectrum simultaneously by integrating an on-chip beam splitter in the output of the interferometer, where one path can be directed to a silicon detector and the other path to an InGaAs one. The measured field autocorrelation signal using an InGaAs detector with our device is shown in Fig. 5 along with the corresponding calculated pulse optical spectrum. The calculated optical spectrum is compared to the optical spectrum measured using an optical spectrum analyzer (OSA) and a good match is found given the higher resolution of the OSA measurement.

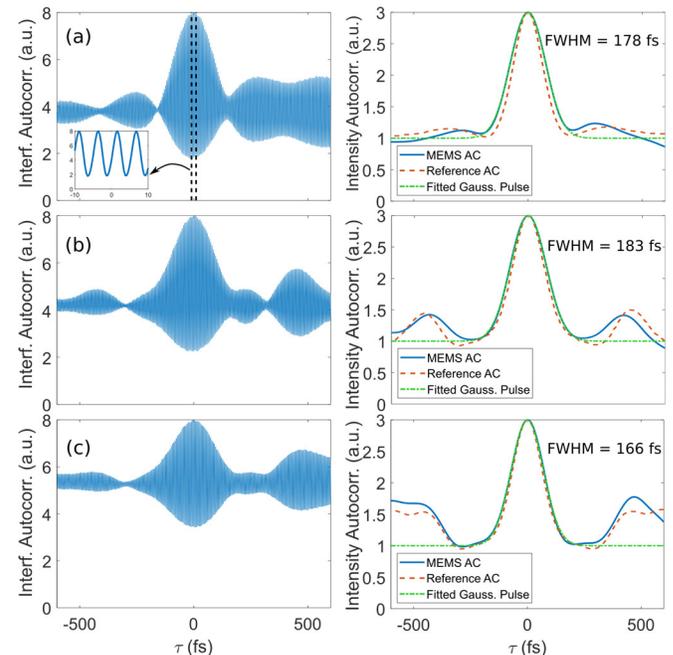

Fig. 4. Measured interferometric autocorrelation (left) and intensity autocorrelation (right) using the proposed autocorrelator for the mode-locked laser pump power of (a) 180 mW, (b) 215 mW and (c) 240 mW. The intensity AC is compared to that of a reference commercial autocorrelator. A Gaussian pulse fitting for the measured autocorrelation is also plotted as the dash-dotted curve. The inset of the top-left figure shows a zoom in of the interferometric AC.

TABLE I
MEASURED PULSE WIDTH USING THE PROPOSED DEVICE AND USING A REFERENCE COMMERCIAL DEVICE

| Pump Power (mW) | Proposed Device Pulse Width (fs) | Reference Device Pulse Width (fs) |
| --- | --- | --- |
| 180 | 126 | 103 |
| 215 | 130 | 116 |
| 240 | 117 | 108 |

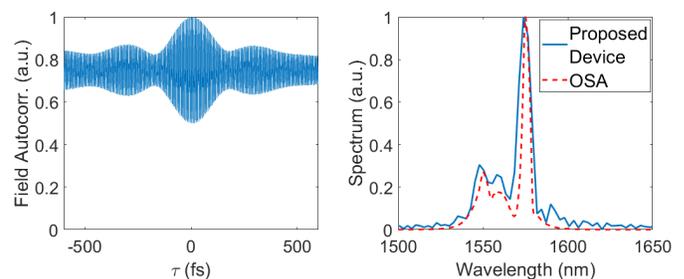

Fig. 5. Measured field autocorrelation signal (left) and the corresponding pulse optical spectrum (right). The optical spectrum is compared to the measurement of an OSA.

## C. Device Limitations

The autocorrelator scanning range is limited by the moving mirror travel range, which is about 200 µm in our case. This corresponds to a scanning range of about 1.2 ps. The autocorrelator sensitivity is defined as the minimum detected product of the peak and average power [6]. It was measured by adding a variable optical attenuator after the mode-locked fiber laser. The attenuation was increased until the minimum detectable pulse was reached. The sensitivity was then calculated to be about 8 $W^2$. The relatively low sensitivity of the device is due to the insertion loss of the MEMS-based interferometer. The minimum measurable pulse width is limited by the silicon dispersion, due to the relatively high silicon group velocity dispersion (GVD), which is equal to 1108 $fs^2$/mm at 1560 nm. The silicon propagation distance for MEMS-based Michelson interferometers ranges from 500 μm to 1000 μm. Assuming an acceptable error of 10%, the minimum measurable pulse width can be estimated to be equal to 58 fs and 82 fs for a silicon propagation distance of 500 μm and 1000 μm, respectively.

## IV. SIMULATION RESULTS

The measured autocorrelation signals have some discrepancies from the ideal theoretical case due to different effects in the MEMS interferometer as will be explained in this section. Asymmetry could be observed in the measured interferometric autocorrelation. Also, the measured intensity and interferometric autocorrelation signals were found to be not following the theoretical ratio between the maximum value and the background level. For the sake of comparison with the reference autocorrelator measurements, the intensity autocorrelations measured by the MEMS-based autocorrelator shown in Fig. 4 were scaled and down-shifted. The unscaled intensity autocorrelation measured at a pump power of 215 mW is shown in Fig. 6(a).

Propagation of the beam in the MEMS-based Michelson interferometer has some effects on the input light beam such as divergence and dispersion. Moreover, the interferometer is not ideal as the mirrors may be slightly tilted from the ideal position, and the amplitude of the two interfering signals may be different. A simulation model is developed as discussed in the following subsections to study these effects and their impact on the measured autocorrelation signals.

### A. Divergence Effect

The input light beam to the interferometer is not perfectly collimated and, hence, suffers from beam divergence. Due to the unequal distance travelled by the two interfering beams at non-zero optical path difference values, the two beams have different values of width and a different phase profile.

A simulation model based on the Gaussian beam propagation is developed to study this effect. The two beams are assumed to be initially identical Gaussian beams with a beam waist radius of $W_0$. The profiles of the beams are then calculated after propagation in the interferometer. The total field at each point on the detector head is calculated. Then, the autocorrelation signal is calculated for both the field autocorrelation (without non-linearity), and the interferometric autocorrelation, by changing the temporal delay ($\tau$) between the two pulses and calculating the detector current at each delay value. The intensity autocorrelation (fringe-averaged) is also calculated from the interferometric autocorrelation.

Two temporal pulse shapes $E_{pulse_t}(t)$ are chosen in our simulation model. The first one has a symmetric intensity profile to have side lobes like the practically measured pulse (Fig. 6(b)). The input pulse width is 240 fs. The second pulse has an asymmetric pulse shape, with the intensity profile shown in Fig. 6(b). Since the divergence of Gaussian beams is wavelength-dependent, the input pulse electric field in time domain $E_{pulse_t}(t)$ is transformed to the frequency $f$ domain using the Fourier transform.

$$E_{pulse_f}(f) = F.T.\{E_{pulse_t}(t)\} \quad (5)$$

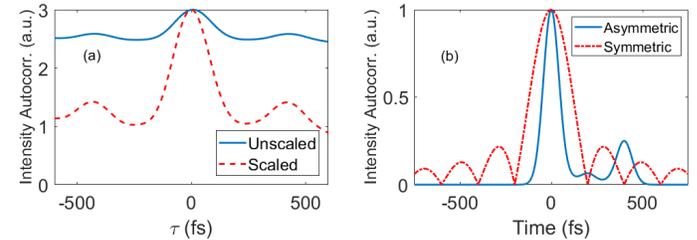

Fig. 6. (a) Intensity autocorrelation measured at pump power 215 mW before and after shifting and scaling. (b) The pulse intensity profile for the asymmetric pulse used in simulation.

The total field at the detector head $E_{det_f}(f, x, y; \tau)$ at each delay is calculated in the frequency (or wavelength) domain as the sum of the two interfering Gaussian electric fields $E_1$ and $E_2$ such that:

$$E_{det_f}(f, x, y; \tau) = E_1(f, x, y; z_1) + E_2(f, x, y; z_2) \quad (6)$$

where $x$ and $y$ are the transverse space coordinates, $z_1$ is the distance traveled by the first beam, $z_2 = z_1 + \tau c$ is the distance traveled by the second beam and $c$ is the speed of light in air. The fields $E_1$ and $E_2$ are defined as [22]:

$$E_i(f, x, y; z_i) = E_{pulse_f}(f) \frac{W_0}{W(z_i, f)} \exp\left(-\frac{x^2 + y^2}{W^2(z_i, f)}\right)$$
$$\cdot \exp\left(-j\left(k(f)z_i + \frac{k(f)(x^2+y^2)}{2R(z_i,f)} - \xi(z_i, f)\right)\right) \quad (7)$$

where $i = 1$ and 2, $W_0$ is the input beam waist radius, $W(z_i, f)$ is the beam radius at distance $z_i$ from the beam waist, $R(z_i, f)$ the radius of curvature of the beam wavefront at $z_i$, $\xi(z_i, f)$ is the Gouy phase [22] at $z_i$ and $k(f) = 2\pi f/c$ is the propagation constant.

Due to the non-linearity of the detector (as TPA is considered the main mechanism for non-linearity in calculating the interferometric autocorrelation), the output current from the detector is dependent on both the interfering pulses temporal profile and their average value. Therefore, the detector output current should be calculated at each time point of the interfering pulses at each value of delay. Thus, the total field at the detector is calculated in the time domain by calculating the inverse Fourier transform of $E_{det_f}(f, x, y; \tau)$:

$$E_{det_t}(t, x, y; \tau) = I.F.T.\{E_{det_f}(f, x, y; \tau)\}. \quad (8)$$

The detector output current $I_{interf}(\tau)$ is calculated for the interferometric autocorrelation case as a function of delay assuming a non-linear detector that employs TPA. It is given as:

$$I_{interf}(\tau) = \beta \int \int \int |E_{det_t}(t,x,y;\tau)|^4 dt\, dx\, dy \quad (9)$$

Also, the field autocorrelation $I_{field}(\tau)$ assuming a linear detector is calculated as:

$$I_{field}(\tau) = \alpha \int \int \int |E_{det_t}(t,x,y;\tau)|^2 dt\, dx\, dy \quad (10)$$

where $\alpha$ and $\beta$ are constants that depend on the responsivity of the detector and its two-photon absorption coefficient, respectively. Their values are not important as the autocorrelation signals are normalized to their maximum value. Fig. 7 shows the simulated interferometric and field autocorrelation signals at $W_0$ values of 5 µm, 10 µm and 20 µm, as well as for the asymmetric pulse case at a $W_0$ of 10 µm. The amplitude of the autocorrelation side lobes varies by changing $W_0$ and a slight asymmetry can be observed in the resulting autocorrelation for both the interferometric and field autocorrelation signals.

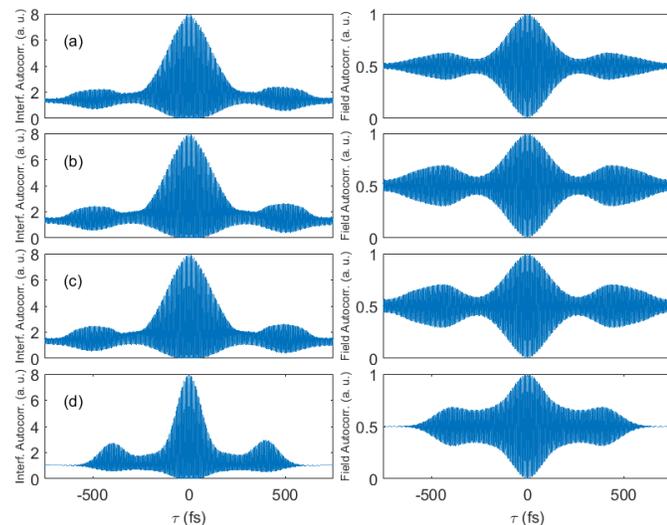

Fig. 7. Simulated interferometric autocorrelation (left) and field autocorrelation (right) signals for $W_0$ values of (a) 5 µm, (b) 10 µm, and (c) 20 µm; and (d) for the asymmetric pulse at a $W_0$ of 10 µm.

### B. Dispersion Effect

The fabricated interferometer uses a silicon beam splitter. Silicon has a relatively large GVD of 1108 fs$^2$/mm at 1560 nm, as calculated from the Sellmeier coefficients of silicon [23]. The fabricated interferometer structure is compensated for dispersion in the sense that both interferometer arms have the same silicon propagation distance [24]. However, this dispersion compensation may not be perfect and some silicon propagation distance mismatch may be present in the fabricated interferometer. To simulate this effect, the second interfering beam $E_2(f,x,y;z_2)$ is given an extra phase term of $\exp(-jn_{si}(f)\,k(f)\,\Delta d)$, where $n_{si}$ is the silicon refractive index calculated using the Sellmeier coefficients of silicon as a function of frequency and $\Delta d$ is the silicon propagation distance mismatch. The simulated interferometric as well as field autocorrelation signals are shown in Fig. 8(a), 8(b) and 8(c) for $\Delta d$ values of 50 µm, 200 µm, 600 µm, respectively. Fig. 8(d) shows the interferometric and field autocorrelation signals for the asymmetric pulse at $\Delta d$ of 200 µm. It is observed that introducing a dispersion mismatch between the arms affects the interference peak-to-background ratio for the interferometric autocorrelation and the interference visibility for the field autocorrelation. A clear asymmetry can also be observed for the interferometric autocorrelation of the asymmetric pulse, but not for the field autocorrelation. The intensity autocorrelation is shown in Fig. 9(a), which shows that the intensity autocorrelation is much less affected by the mismatch, and shows no asymmetry for an asymmetric pulse shape.

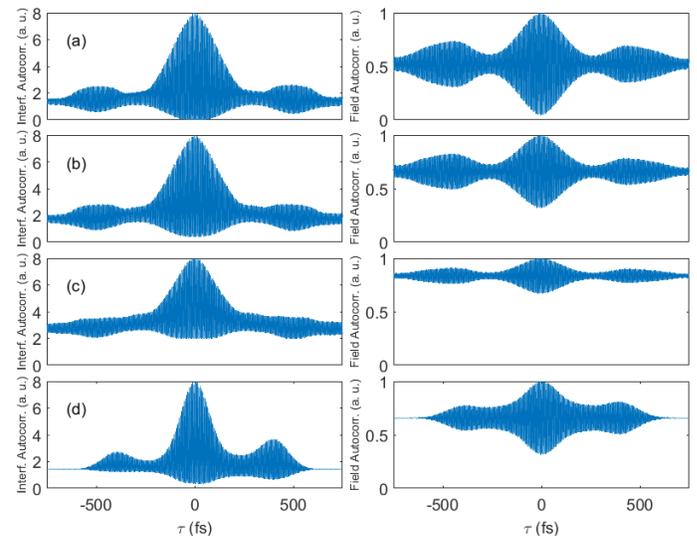

Fig. 8. Simulated interferometric autocorrelation (left) and field autocorrelation (right) signals for a silicon propagation distance mismatch $\Delta d$ of (a) 50 µm, (b) 200 µm, and (c) 600 µm; and (d) for the asymmetric pulse at a $\Delta d$ of 200 µm.

### C. Surfaces Tilt Angle Effect

The fabricated MEMS interferometer beam splitter and reflecting mirrors are not perfectly vertical due to fabrication technology limitations and, hence, the interfering beams may have slightly shifted centers at the detector [25]. This effect is introduced to the simulation model by making one of the beams propagate along a slightly tilted axis from the $z$ direction. The simulated interferometric and field autocorrelation signals are shown in Fig. 10(a) and 10(b) for tilt angles of $0.1^0$ and $0.5^0$ respectively, while Fig. 10(c) shows them for the asymmetric pulse case at a tilt angle of $0.5^0$. The tilt angle affects the pulse side lobe shape in addition to affecting the interference visibility. The field interference visibility becomes less than unity and the interferometric autocorrelation peak to background ratio is no longer 8 to 1. Also, the interferometric autocorrelation becomes slightly asymmetric. However, the intensity autocorrelation is much less affected by the tilt angle and it shows no noticeable change in shape by changing the tilt angle to from $0.1^0$ to $0.5^0$ as shown in Fig. 11.

### D. Unequal Amplitude of Interfering Beams Effect

The two interfering beams travel different paths after splitting and hence may be subject to unequal losses. This may also happen due to the different spot sizes at the detector resulting from different propagation distance, which may cause the larger spot to be truncated due to the limited size of the detector. To study this effect, a simple model is developed for the interferometer, where the two beams are assumed to have unequal amplitude. The amplitude ratio between the electric field of the two beams was varied from 1 to 0.2.

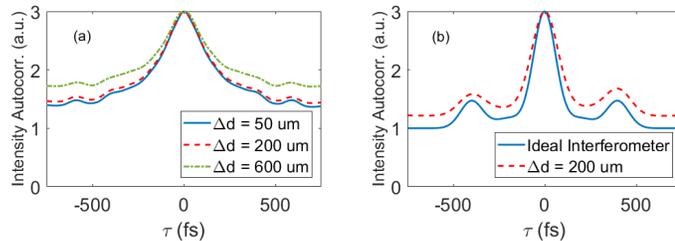

Fig. 9. (a) Simulated intensity autocorrelation signals for a silicon propagation distance mismatch $\Delta d$ of 50 µm, 200 µm and 600 µm. (b) The intensity autocorrelation signal for the asymmetric pulse at $\Delta d$ of 200 µm compared to an ideal interferometer.3

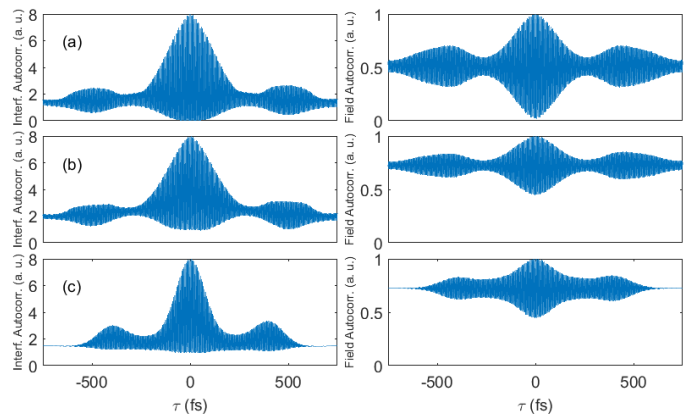

Fig. 10. Simulated interferometric autocorrelation (left) and field autocorrelation (right) signals for surface tilt angle values of (a) $0.1^0$ and (b) $0.5^0$; and (c) for the asymmetric pulse at a tilt angle of $0.5^0$.

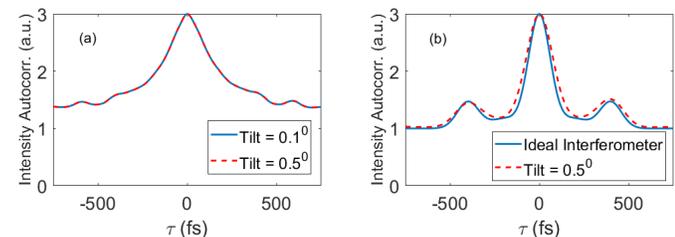

Fig. 11. (a) Simulated intensity autocorrelation signals for surface tilt angle values of (a) $0.1^0$ and (b) $0.5^0$. (b) The intensity autocorrelation signal for the asymmetric pulse at a tilt angle of $0.5^0$ compared to an ideal interferometer.

The simulated results for interferometric and field autocorrelation signals are shown Fig. 12(a), 12(b) and 12(c) for amplitude ratios of 1, 0.5 and 0.25, respectively. Fig. 12(d) shows the interferometric and field autocorrelation signals for the asymmetric input pulse. Fig. 13(a) shows the intensity autocorrelation signals for the symmetric pulse at different values of pulse amplitude ratio, while Fig. 13(b) shows the intensity autocorrelation signal for the asymmetric pulse at a pulse amplitude ratio of 0.2 compared to the ideal interferometer case. The amplitude ratio is shown to have only a scaling effect on the field and intensity autocorrelation signals (a change in peak to background ratio) and no effect at all on their shape. However, for the interferometric autocorrelation, the amplitude ratio also affects the shape of the side lobes of the autocorrelation signal in addition to its peak to background ratio. In addition, for the asymmetric input pulse shape, a clear asymmetry is observed.

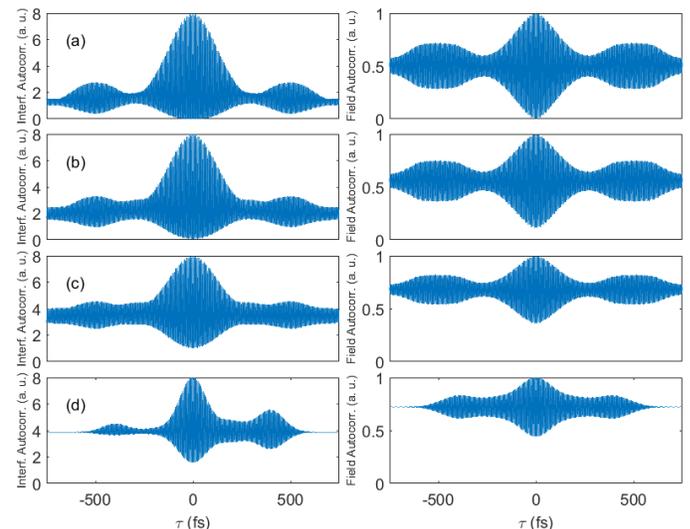

Fig. 12. Simulated interferometric autocorrelation (left) and field autocorrelation (right) signals for interfering pulses amplitude ratio of (a) 1, (b) 0.5, (c) 0.25; and (d) for the asymmetric pulse at a pulse amplitude ratio of 0.2

The asymmetry in the interferometric autocorrelation can be attributed to the second integral in (3), which for a small delayed pulse becomes a cross correlation between $|E_p(t)|^2 E_p(t)$ and $E_p(t)$. Hence, it is asymmetric if $E_p(t)$ is not symmetric. The autocorrelation signals in Fig. 12(d) and Fig. 13(b) show a good agreement with the measurement results in Fig. 4 that indicates that the measured pulse was an asymmetric pulse, which is a known case in mode-locked fiber lasers [26], [27].

## V. Conclusion

A compact MEMS-based optical autocorrelator has been presented, which can measure both the pulse width and the optical power spectral density. The device operates in the wavelength range of 1100-2000 nm and has a scanning range of 1.2 ps. Different femtosecond pulses were measured and showed to have a good agreement with the measurements done by a commercial autocorrelator. A simulation model has been presented to study the effects of the light divergence, the beam splitter dispersion, the surfaces tilt angle and the amplitude mismatch of interfering beams that may exist in our system due to the use of a MEMS-based interferometer instead of a conventional one. Simulation results show impairments in the interferometric autocorrelation signals similar to those in the measurement results. Also, the intensity autocorrelation signals illustrated in both the experimental and simulation results are shown to have higher immunity to the light divergence, the beam splitter dispersion, and the surfaces tilt angle than the interferometric autocorrelation signals.

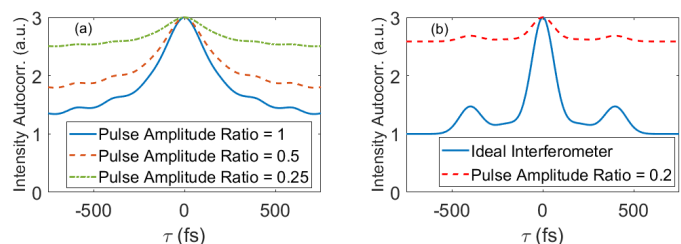

Fig. 13. (a) Simulated intensity autocorrelation signals for different values of interfering pulses amplitude ratio. (b) The intensity autocorrelation signal for the asymmetric pulse at a pulse amplitude ratio of 0.2 compared to an ideal interferometer.